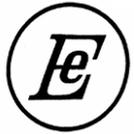

**17th INTERNATIONAL SYMPOSIUM on POWER ELECTRONICS - Ee 2013**

NOVI SAD, SERBIA, October 30th – November 1st, 2013

# SIMULATION OF SWITCHING CONVERTERS ON THE LEVEL OF AVERAGED VOLTAGES AND CURRENTS

**Aleksandra Lekić, Predrag Pejović**
University of Belgrade, Faculty of Electrical Engineering, Belgrade, Serbia

**Abstract:** *An algorithm for simulation of switching converters is proposed in the paper. The algorithm is based on simulation of averaged circuit model applying "switching cell" concept, and construction of instantaneous values of the waveforms using quasi steady state and linear ripple approximation. Simulation covers converters operating both in the continuous and the discontinuous conduction mode. Application of the algorithm is demonstrated by simulation results of all three of the basic converters: buck, boost and buck-boost, as well as a flyback converter, which required slight generalization of the switching cell concept.*

## 1. INTRODUCTION

Motivation for this research lies in the fact that the most of available circuit simulators is general-purpose oriented [6] and apply unnecessarily complex device models, characterized by continuous and differentiable functions. On the other hand, the most important components of power electronics are switches, characterized by discontinuities in their constructive equations. Attempts to simulate power electronics circuits applying general-purpose simulation tools frequently yields convergence problems [7], reducing effectiveness of simulation in the design process, especially in cases when the converter control is focused. This motivated development of specialized simulation tools, intended to be applied in simulation of long term transients in power electronics circuits [7, 8, 9].

To avoid convergence problems, the approach proposed here is based on simulation of averaged circuit model [1, 2], with averaging applied on the "switching cell" level [3, 4, 5], and construction of actual waveforms, i.e. instantaneous values of the waveforms, applying quasi steady state approximation [10] and linear ripple approximation [1]. The algorithm presents an extension of the concepts proposed in [10] to cover discontinuous conduction mode and to generalize formulation of equations. Proposed approach results in a fast algorithm that provides instantaneous values of voltages and currents and their average values, minima and maxima. Obtained waveforms could be post-processed further, aiming spectra of the waveforms, their rms values, or any other waveform parameter.

## 2. APPLIED SWITCHING CELL

In this section, switching cell, used to describe both the CCM and DCM operating modes, is presented. Switching cells are treated as a three terminal devices. In this paper, two families of the switching cells are described: for all three of the basic converters (buck, boost, and buck-boost), switching cells classified as cell A family in [3] are used; for flyback converter, slightly modified switching cell that includes a transformer is used.

Two types of switching cells are considered for each family: synchronous one, that applies two bidirectional switches ( $\text{state}(S2) = \neg \text{state}(S1)$ ), as well as a common type that applies a controlled switch and a diode. The converter with synchronous rectification remains in continuous conduction mode (CCM), while the converter with the diode might enter discontinuous conduction mode (DCM). The switching cells are summarized in Table 1.

Averaged model of the switching cell used in simulations is shown in the Fig. 1. This model presents switching cell operating both in continuous and in discontinuous conduction mode, and preserves the circuit topological structure.

Table 1. *Types of switching cells.*

| | with bidirectional switches and the inductor | with switch, diode and inductor |
|---|---|---|
| basic cell | 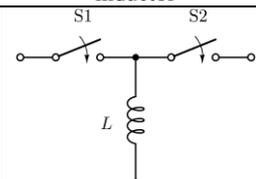 | 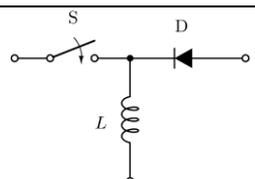 |
| flyback cell | 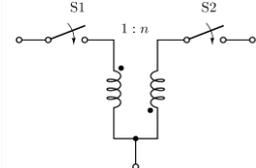 | 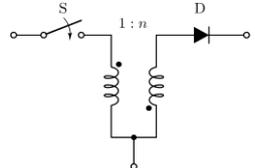 |

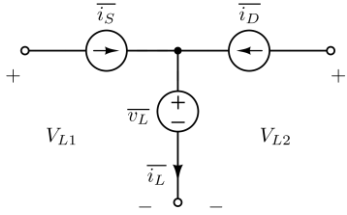

Fig. 1. *Averaged model of a switching cell.*

Focusing to a switching interval $0 \leq t < T_S$, $T_S = 1/f_S$ and assuming that $i_{L0}$, $i_{L1}$ and $i_{L2}$ are inductor currents at the beginning of the period: $i_{L0} = i_L(0)$, at the moment $t = dT_S$: $i_{L1} = i_L(dT_S)$ and at the end of the switching period: $i_{L2} = i_L(T_S)$, using linear ripple approximation and quasi steady state approximation (as proposed in [10]), the set of the equations describing basic switching cell given in Table 1 is formulated.

According to typical waveforms of $i_S$, $i_D$, and $v_L$ over a switching period, for both of the conduction modes following expressions for averages of currents of switching devices and the inductor voltage are derived in the form:

$$i_{L1} = i_{L0} + \frac{V_{L1}}{L} d T_S \quad (1)$$

$$i_{L2} = i_{L1} + \frac{V_{L2}}{L} d_p T_S \quad (2)$$

$$\overline{v_L} = d V_{L1} + d_p V_{L2} \quad (3)$$

$$\overline{i_S} = \frac{d}{2}(i_{L0} + i_{L1}) \quad (4)$$

$$\overline{i_D} = \frac{d_p}{2}(i_{L1} + i_{L2}) \quad (5)$$

where $d_p$ equals $d_p = 1 - d$ in CCM and $d_p = d_2$ in DCM, where $d_2$ is

$$d_2 = -\frac{V_{L1}}{V_{L2}} d \quad (6)$$

and the converter operates in the CCM if $d + d_2 \geq 1$ and in the DCM if $d + d_2 < 1$. The average model of the switching cell consists of (1)–(6) and covers both of the conduction modes according to the accompanying inequalities.

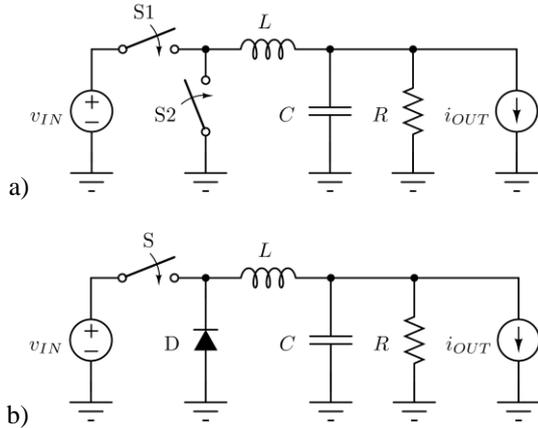

a)

b)

Fig. 2. *The buck converter: a) with synchronized bidirectional switches; b) with a switch and a diode.*

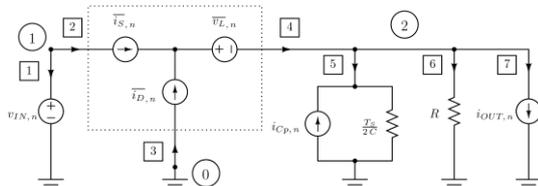

Fig. 3. *Discretized averaged circuit model of the buck converter.*

Switching cell for flyback converter, presented in Table 1, is described by a slightly modified set of equations, due to the presence of transformer. Furthermore, in comparison to original flyback converter, the cell is modified such that galvanic isolation is removed, in order to obtain connected graph which is a requirement to perform numerical simulation, since the numerical simulation can handle only circuits with unique solutions. It is still open question whether this connection should be performed at the switching cell level or at the circuit level after the netlist is formed, which is going to be discussed in future publications.

## 3. APPLICATION OF THE SWITCHING CELL MODEL AND CIRCUIT EQUATIONS FORMULATION

To illustrate application of switching cells, an example of the buck converter, depicted in Fig. 2.a, is considered in this section. This example addresses synchronous buck converter, which always operates in the CCM. Identifying the switching cell of Fig. 1. in the buck converter circuit of Fig. 2.a results in the circuit of Fig. 3, where the capacitor is replaced by its discretized equivalent according to the trapezoidal integration formula. All of the currents and voltages in the circuit shown in Fig. 3 represent averaged values during $n^{th}$ switching period, for $nT_S \leq t < (n+1)T_S$.

Discretized averaged circuit model of Fig. 3 is in each time step solved applying modified nodal analysis technique [6]. Other equations necessary to perform the simulation treat computation of independent sources in the circuit of Fig. 3, and besides the values of originally independent sources ($v_{IN}$ and $i_{OUT}$) arise from the trapezoidal integration formula and the equations that describe the switching cell. In the example of Fig. 3, the set of equations in a slightly optimized form of modified nodal analysis is given by

$$v_{1,n} = v_{IN,n} \quad (7)$$

$$v_{2,n} - v_{C,n} = 0 \quad (8)$$

$$i_{1,n} + \overline{i_{S,n}} = 0 \quad (9)$$

$$i_{C,n} - \overline{i_{L,n}} + \frac{v_{2,n}}{R} = -i_{OUT,n} \quad (10)$$

$$v_{C,n} - \frac{T_S}{2C} i_{C,n} = v_{C,n-1} + \frac{T_S}{2C} i_{C,n-1} \quad (11)$$

$$\overline{i_{S,n}} - \frac{d_n}{2} i_{L1,n} = \frac{d_n}{2} i_{L0,n} \quad (12)$$

$$\overline{i_{D,n}} - \frac{d_{p,n}}{2}(i_{L1,n} + i_{L2,n}) = 0 \quad (13)$$

$$\overline{i_{L,n}} - \overline{i_{S,n}} - \overline{i_{D,n}} = 0 \quad (14)$$

$$\overline{v_{L,n}} - d_n V_{L1,n} - d_{p,n} V_{L2,n} = 0 \quad (15)$$

$$i_{L1,n} - \frac{V_{L1,n}}{L} d_n T_S = i_{L0,n} \quad (16)$$

$$i_{L2,n} - i_{L1,n} - \frac{V_{L2,n}}{L} d_{p,n} T_S = 0 \quad (17)$$

$$V_{L1,n} - v_{1,n} + v_{2,n} = 0 \quad (18)$$

$$V_{L2,n} + v_{2,n} = 0 \quad (19)$$

and this linear system has to be solved in each time step in order to obtain desired waveforms. Although slightly optimized and suited for the purpose of demonstrating the algorithm, the system of 13 equations is large regarding relatively small size of the circuit being simulated. This directs further research into application of sparse matrix techniques (less likely to produce the desired result) or reduction of the system size extracting equations that describe the switching cell (more likely to improve the performance).

## 4. SIMULATION ALGORITHM

Application of the algorithm is demonstrated by developing a simple program that simulates startup transient. From given list of the elements, the program forms matrices $\mathbf{A}$, $\mathbf{x}$ and $\mathbf{z}$ where $\mathbf{A}\mathbf{x} = \mathbf{z}$, according to a formal procedure. Matrix $\mathbf{x}$ is an $(n+m+k) \times 1$ vector which holds the unknown quantities. Top $n$ quantities represent node voltages, $m$ elements in the middle represent currents through voltage sources in the circuit and the bottom $k = 9$ elements are the capacitor voltage and 8 variables of the switching cell. Matrix $\mathbf{A}$ represents coefficients that multiplied by x result in the right and side vector $\mathbf{z}$. In the CCM, matrix $\mathbf{A}$ is constant, but in the DCM some of the elements have to be replaced for each iteration. Right hand side vector $\mathbf{z}$ consists of original independent sources and the independent sources introduced by the discretization and the switching cell model, being dependent from the network solution in the previous iteration.

In the case of the buck converter shown in Fig. 2, with the corresponding model depicted in Fig. 3, described by (7)–(19), the system that describes the circuit in $n^{th}$ switching period is given by

$$\begin{bmatrix} 1 & 0 & 0 & 0 & 0 & 0 & 0 & 0 & 0 & 0 & 0 & 0 \\ 0 & 1 & 0 & 0 & -1 & 0 & 0 & 0 & 0 & 0 & 0 & 0 \\ 0 & 0 & 1 & 0 & 0 & 1 & 0 & 0 & 0 & 0 & 0 & 0 \\ 0 & \frac{1}{R} & 0 & 1 & 0 & 0 & -1 & 0 & 0 & 0 & 0 & 0 \\ 0 & 0 & 0 & -R_C & 1 & 0 & 0 & 0 & 0 & 0 & 0 & 0 \\ 0 & 0 & 0 & 0 & 0 & 1 & 0 & 0 & -\frac{d_n}{2} & 0 & 0 & 0 \\ 0 & 0 & 0 & 0 & 0 & 0 & 1 & 0 & -\frac{d_{p,n}}{2} & -\frac{d_{p,n}}{2} & 0 & 0 \\ 0 & 0 & 0 & 0 & -1 & -1 & 1 & 0 & 0 & 0 & 0 & 0 \\ 0 & 0 & 0 & 0 & 0 & 0 & 0 & 1 & 0 & 0 & -d_n & -d_{p,n} \\ 0 & 0 & 0 & 0 & 0 & 0 & 0 & 1 & 0 & -d_n G_L & 0 & 0 \\ 0 & 0 & 0 & 0 & 0 & 0 & 0 & -1 & 1 & 0 & 0 & -d_{p,n} G_L \\ -1 & 1 & 0 & 0 & 0 & 0 & 0 & 0 & 0 & 1 & 0 & 0 \\ 0 & 1 & 0 & 0 & 0 & 0 & 0 & 0 & 0 & 0 & 0 & 1 \end{bmatrix} \begin{bmatrix} v_{1,n} \\ v_{2,n} \\ i_{IN,n} \\ i_{C,n} \\ v_{C,n} \\ i_{S,n} \\ i_{D,n} \\ i_{L,n} \\ v_{L,n} \\ i_{L1,n} \\ i_{L2,n} \\ V_{L1,n} \\ V_{L2,n} \end{bmatrix} = \begin{bmatrix} v_{IN,n} \\ 0 \\ 0 \\ -i_{OUT,n} \\ v_{C,n-1} + R_C i_{C,n-1} \\ \frac{d_n}{2} i_{L0,n} \\ 0 \\ 0 \\ 0 \\ i_{L0n} \\ 0 \\ 0 \\ 0 \end{bmatrix}$$

(20)

where $R_C = T_S/(2C)$ and $G_L = T_S/L$. Solution of the circuit during $n^{th}$ switching period is

$$\mathbf{x}_n = \mathbf{A}_n^{-1} \mathbf{z}_n. \quad (21)$$

Simulation of the averaged circuit involves the following steps: forming the equation system of (21), which corresponds to updating parameters of $\mathbf{A}_n$ and $\mathbf{z}_n$, and computing of $\mathbf{x}_n$.

While generating the circuit equations, it should be checked in which mode the converter would operate in the next switching period. The condition for resolving this issue is related to the inductor current, and assuming fixed voltages across the switching cell ports during the switching period and knowing the initial value of the inductor current, it can be predicted in advance to the simulation step. This approach limits the step size to correspond to the switching period.

After the values of vector $\mathbf{x}$ are computed in a number of time points in a considered interval, averaged waveforms of the inductor current and the capacitor voltage can be plotted, as a direct result of simulation. However, to construct instantaneous waveforms, quasi steady state approximation is used [10].

Essential waveform to be constructed is the waveform of the inductor current. During each switching interval, inductor current at the beginning, at the moment $t = dT_S$ when switching occurs, and at the end of the interval are known. In the case of the discontinuous conduction mode, an additional point where the inductor current reaches zero is computed. Since the linear ripple approximation is applied, the instantaneous waveform of the inductor current is obtained by plotting these values and connecting them by straight line segments. In a similar manner, instantaneous voltage of the capacitor can be computed superimposing the ripple to the average value.

## 5. SIMULATION EXAMPLES

To illustrate application of the algorithm, two examples are given in this section. The first example is buck converter, which demonstrates simulation results for all of the basic converters. In the second example flyback converter which requires modified switching cell is presented.

### 4.1. Example 1: Buck converter

In this example, two types of the buck converter are used as benchmark circuits: a synchronous buck rectifier that applies switching cell depicted in Fig. 2.a, and a common buck converter that applies a controlled switch and a diode, depicted in Fig. 2.b. For both of the converters, startup transient is simulated. Expected average of the output voltage in the steady state is $V_{OUT} = 5$ V, and the inductor current average value is $I_L = 5$ A, which should be achieved at the end of the simulation.

Simulation results are presented in Fig. 4, where the waveforms in the left column correspond to the converter with synchronous rectification, while the right column corresponds to the converter that applies the diode and might operate in the discontinuous conduction mode. In the diagrams of the inductor current, time intervals in which the converter operates in the DCM can be identified. Waveforms that correspond to the converter that enters the DCM in some time intervals expose faster convergence towards the steady state operation. Contribution of this paper in comparison to [10] is in enabling simulation of switching cells in discontinuous conduction modes.

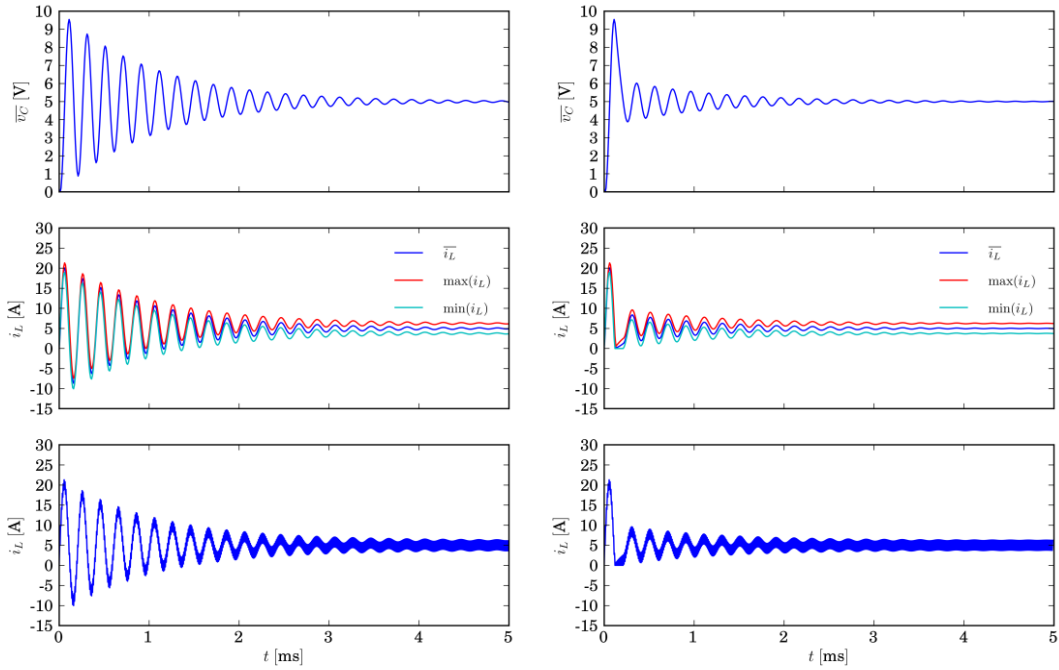

Fig. 4. *Simulation results of the example 1 : converter of Fig. 2.a, left column; converter of Fig. 2.b, right column. Parameters:* $R = 5\,\Omega$, $I_{OUT} = 4\,\text{A}$, $L = 10\,\mu\text{H}$, $C = 100\,\mu\text{F}$, $V_{IN} = 10\,\text{V}$, $f_S = 100\,\text{kHz}$, $D = 0.5$.

### 4.2. Example 2: Flyback converter

As a next example, consider the flyback converter shown in Fig. 5.a. as an example of the synchronous flyback rectifier and the common flyback converter depicted in Fig. 5.b. Simulation results are presented in Fig. 6. Expected average of the output voltage in the steady state is $V_{OUT} = 20\,\text{V}$, and the inductor current average value at the end of simulation should be $I_L = 20\,\text{A}$, which actually is achieved in the simulation.

Diagrams are constructed for the time interval of 5 ms. Since the switching frequency is 100 kHz, the waveforms are obtained for 500 switching periods, i.e. 500 samples over the given time interval. The waveforms of $i_L$ and $v_C$ are in agreement with the theoretical predictions. This demonstrates that the basic switching cell can be modified according to Table 1 to model isolated converters such as flyback.

The simulation results are obtained by a program written specifically for this purpose, in Python programming language, in Pylab environment, relying entirely on free software tools. The Pylab environment provides access to matplotlib package, which was used to plot the graphs shown in Figs. 4 and 6. The numerical simulation without generating the plots was completed in 50 ms on a PC computer equipped with Intel Core 2 Duo P8400 processor run at 2.26 GHz under Ubuntu 13.04 operating system. It should be taken into account that the algorithm is implemented in Python programming language, which is interpretative, thus implementation in C should run even faster.

The program requires a text file that contains list of elements (capacitors, resistors, switching cells, etc.) and for a given duty ratio, switching frequency, transient duration and initial conditions, generates matrices **A**, **x** and **z** and plots the inductor current and the capacitor voltage waveforms in the form given in Figs. 4 and 6. Switching cells specified by Table 1 are specified as three terminal elements.

### 4. CONCLUSION

An accurate and efficient simulation algorithm for switching converters is proposed in this paper. The algorithm performs simulation on the averaged circuit level, and applies switching cell concept. Convergence problems, as well as problems of determination of state of piecewise linear elements are avoided by the use of the switching cell concept, linear ripple approximation and quasi steady state approximation. As a result, fast algorithm for averaged circuit simulation is obtained with the possibility to reconstruct waveforms of actual circuit currents and voltages. Both the continuous and the discontinuous conduction mode can be handled by the proposed method, primarily due to the application of the switching cell concept. Application of the switching cell concept is in increment in comparison to [10], and enables simulation of converters in the discontinuous conduction mode.

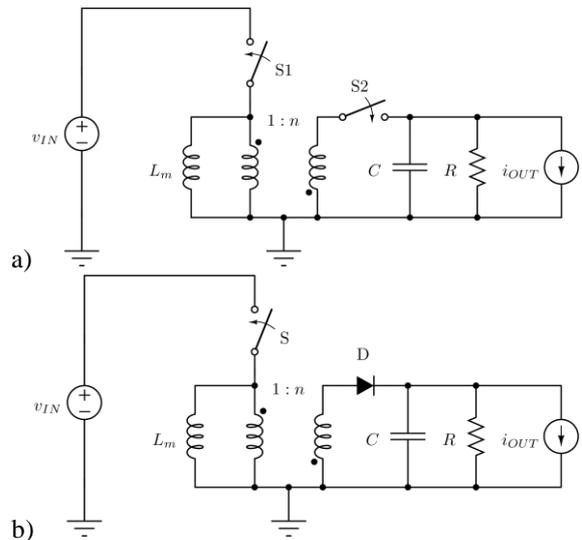

Fig. 5. *Simulated flyback converters, a) with bidirectional switches and the inductor and b) with switch, diode and inductor.*

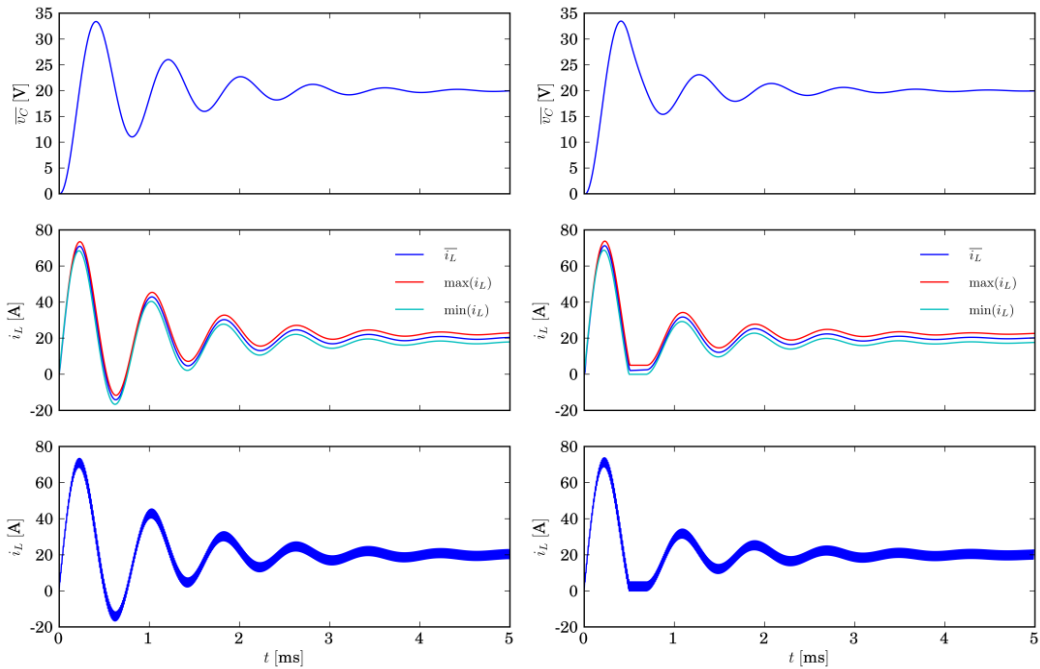

Fig. 6. *Simulation results of the example 2: converter of Fig. 5.a, left column; converter of Fig. 5.b, right column. Parameters:* $R = 5\ \Omega$, $I_{OUT} = 1\ \text{A}$, $L_m = 10\ \mu\text{H}$, $C = 100\ \mu\text{F}$, $V_{IN} = 10\ \text{V}$, $f_S = 100\ \text{kHz}$, $D = 0.5$, $n = 2$.

The algorithm discussed above is implemented in a program written in Python programming language. Regardless the fact that Python is an interpretive programming language, simulation is really fast, taking only 50 ms. This indicates that proposed algorithm is promising to become a useful and efficient tool for power electronic circuits. Further development is directed toward formalizing forming of the equation system and reducing its size, formalizing construction of the instantaneous waveforms on the basis of obtained averaged waveforms, discussion of accuracy and the switching size, and application of the method for true nonlinear circuits, like the converters with the feedback control system applied.